# Origin of the Metallicity Distribution in the Thick Disc

M. S. Miranda[1,2], K. Pilkington[1], B. K. Gibson[3], C. B. Brook[2,1], P. Sánchez-Blázquez[2], I. Minchev[4], C. G. Few[5,3], R. Smith[6,7,8], R. Domínguez-Tenreiro[2], A. Obreja[2], J. Bailin[9,10], and G. S. Stinson[11]

[1] Jeremiah Horrocks Institute, University of Central Lancashire, Preston, PR1 2HE, UK
e-mail: msancho@uclan.ac.uk
[2] Departamento de Física Teórica, Universidad Autónoma de Madrid, Cantoblanco, Madrid, E28049, Spain
[3] E.A. Milne Centre for Astrophysics, Dept of Physics & Mathematics, University of Hull, Hull, HU6 7RX, UK
[4] Leibniz-Institut für Astrophysik Potsdam (AIP), An der Sternwarte 16, D-14482, Potsdam, Germany
[5] School of Physics, University of Exeter, Exeter, EX4 4QL, UK
[6] Yonsei University, Graduate School of Earth System Sciences-Astronomy-Atmospheric Sciences, Yonsei-ro 50, Seoul 120-749, Republic of Korea
[7] Laboratoire AIM Paris-Saclay, CEA/IRFU/SAp, Université Paris Diderot, 91191 Gif-sur-Yvette Cedex, France
[8] Departamento de Astronomía, Universidad de Concepción, Casilla 160-C, Concepción, Chile
[9] Department of Physics & Astronomy, University of Alabama, Tuscaloosa, AL, 35487-0324, USA
[10] National Radio Astronomy Observatory, P.O. Box 2, Green Bank, WV, 24944, USA
[11] Max-Planck-Institut für Astronomie, Königstuhl 17, Heidelberg, 69117, Germany

Monday 14th December, 2015

**ABSTRACT**

*Aims.* Using a suite of cosmological chemodynamical disc galaxy simulations, we assess how (a) radial metallicity gradients evolve with scaleheight; (b) the vertical metallicity gradients change through the thick disc; and (c) the vertical gradient of the stellar rotation velocity varies through the disc. We compare with the Milky Way to search for analogous trends.
*Methods.* We analyse five simulated spiral galaxies with masses comparable to the Milky Way. The simulations span a range of star formation and energy feedback strengths and prescriptions, particle- and grid-based hydrodynamical implementations, as well as initial conditions/assembly history. Disc stars are identified initially via kinematic decomposition, with *a posteriori* spatial cuts providing the final sample from which radial and vertical gradients are inferred.
*Results.* Consistently, we find that the steeper, negative, radial metallicity gradients seen in the mid-plane flatten with increasing height away from the plane. In simulations with stronger (and/or more spatially-extended) feedback, the negative radial gradients invert, becoming positive for heights in excess of ∼1 kpc. Such behaviour is consistent with that inferred from recent observations. Our measurements of the vertical metallicity gradients show no clear correlation with galactocentric radius, and are in good agreement with those observed in the Milky Way's thick disc (locally). Each of the simulations presents a decline in rotational velocity with increasing height from the mid-plane, albeit the majority have shallower kinematic gradients than that of the Milky Way.
*Conclusions.* Simulations employing stronger/more extended feedback prescriptions possess radial and vertical metallicity and kinematic gradients more in line with recent observations. The inverted, positive, radial metallicity gradients seen in the simulated thick stellar discs originate from a population of younger, more metal-rich, stars formed *in-situ*, superimposed upon a background population of older migrators from the inner disc; the contrast provided by the former increases radially, due to the inside-out growth of the disc. A similar behaviour may be responsible for the same flattening seen in the radial gradients with scaleheight in the Milky Way.

**Key words.** galaxies: abundances – galaxies: evolution – galaxies: formation – Galaxy: disc

## 1. Introduction

The classical picture of the Milky Way disc has been one of a two-component structure with a thin disc enshrouded by a thicker stellar disc. Identified by Gilmore & Reid (1983), the thick disc has been the subject of much recent controversy. It was originally thought to contain a distinct/discrete population of stars relative to those of the thin disc, whether divided by luminosity (e.g. Yoachim & Dalcanton 2006), kinematics (e.g. Pasetto et al. 2012) or metallicity (e.g. Lee et al. 2011). Such a 'discrete' thick disc picture is consistent with evidence provided for some external disc galaxies (e.g. Yoachim & Dalcanton 2006; Comerón et al. 2011; Freeman 2012). This picture has been called into question by multivariate mixture models (e.g. Nemec & Nemec 1993) and, more recently, by Bovy et al. (2012), who claim a single, continuous, disc is in better agreement with observations.

Whether the stars seen well above the mid-plane (and still co-rotating with the canonical thin disc) should be thought of as part of a discrete or a semi-continuous structure, we will refer to them colloquially as 'thick disc' stars.

The origin of these thick disc stars remains a primary topic of debate for galactic structure. Seven of the most popular scenarios can be categorised as:

- Brook et al. (2004) suggest that discs are formed thick during the intense gas-rich merger phase at high-redshift; this scenario is supported by observations such as those of Gilmore et al. (2002), Wyse et al. (2006) and Comerón et al. (2011).
- Abadi et al. (2003b) postulate that the thick disc formed from the direct accretion of debris from a now-disrupted SMC-mass satellite; such a satellite mass is required to give the correct stellar metallicities (Freeman 2012).
- Quinn & Binney (1992), Kazantzidis et al. (2008), Villalobos & Helmi (2008) and Qu et al. (2011) favour a scenario in which the thick disc originated from kinematic heating of a pre-existing thin disc.
- Schönrich & Binney (2009) and Loebman et al. (2011) propose that the thick disc might have formed from the radial migration of inner disc stars to the outer regions (but cf. Minchev et al. 2012; Vera-Ciro et al. 2014).
- Kroupa (2002) and Assmann et al. (2011) suggest that the thick disc originated via the 'popping' of star clusters.
- Bournaud et al. (2009) introduce a scenario in which massive clumps scatter stars to high velocity dispersions and form thick discs.
- According to Haywood et al. (2013) the thick disc formed through the birth of stars in a gas layer made thick by turbulence.

The kinematics of the Milky Way thick disc has attracted much attention, no doubt due, in part, to the extraordinary wealth of information to be provided shortly by the Gaia mission (e.g. Rix & Bovy 2013). Ahead of the Gaia Data Releases, the exploitation of extant datasets is both timely and essential for shaping the rapid analysis and dissemination of Gaia's data. Recent, important, efforts in this area include that of Pasetto et al. (2012), who, using data from RAVE (Steinmetz et al. 2006), suggest the thick and thin discs are discrete and separable using stellar kinematics. Similar studies have been performed using data from SDSS (e.g. Carollo et al. 2010), obtaining analogous results. Alternative ways in which to probe and/or isolate thick and thin discs include those of 'chemistry' (e.g. Navarro et al. 2011) and 'distances' (e.g. Carrell et al. 2012, who select stars spatially using only dwarf stars).

In what follows, we examine how the velocity of the stars associated with the disc changes as a function of height above the plane. Observationally, vertical gradients in the rotational velocity of disc stars have been found (e.g. Bond et al. 2010; Casetti-Dinescu et al. 2011; Bovy & Tremaine 2012), with Moni Bidin et al. (2012) claiming a gradient of $dV_\phi/d|z| \approx -25-30$ km s$^{-1}$ kpc$^{-1}$. By comparison with our simulations, these gradients allow us to probe the nature of kinematic transition from thin to thick disc (e.g. is it discrete or continuous?).

The metallicity of the thick disc has been well studied within the Milky Way (e.g. Bensby et al. 2003; Reddy et al. 2006; Ivezić et al. 2008). The spatial variations of the metallicity allow us to test galaxy formation and evolution scenarios. Metallicity gradients within the Milky Way have been studied since Shaver et al. (1983) recognised that the metals were not distributed homogeneously. Since then, radial (e.g. Simpson et al. 1995; Afflerbach et al. 1997), vertical (e.g. Marsakov & Borkova 2005; Soubiran et al. 2008) and azimuthal (e.g. Luck et al. 2011) gradients have been studied extensively in the Milky Way.

In the thin disc of late-type spirals, including the Milky Way, radial metallicity gradients (whether measured in the gas-phase, or young stellar probes) are typically $-0.05$ dex kpc$^{-1}$ (decreasing outwards through the thin disc). Moving away from the mid-plane, into the thick disc ($\sim$1-3 kpc from the mid-plane), the gradient progressively flattens (Cheng et al. 2012) and, indeed, eventually inverts (increasing outwards through the thick disc: Carrell et al. 2012; Anders et al. 2014). Such inversions of the radial metallicity gradient in the thick disc have also been seen in the chemodynamical simulations of Rahimi et al. (2013) and Minchev et al. (2014).

Boeche et al. (2014) have shown recently (using red giant branch stars) how the radial metallicity gradient of the Galaxy changes as a function of height $|z|$ above the plane. Only a small region in $|z|$ is covered by the RAVE sample employed, and so the results are mainly for stars with $|z|<1$ kpc, in the radial range 4.5 to 9.5 kpc. In Boeche et al. (2013), the team also determined radial gradients from the dwarf stars in the Geneva Copenhagen Survey (GCS: Nordström et al. 2004).

Similarly, many studies have checked the vertical metallicity gradient in the Milky Way disc. Kordopatis et al. (2013) have inferred a vertical gradient of $-0.22\pm0.17$ dex kpc$^{-1}$ along the line-of-sight to Sculptor, which corresponds roughly to the vertical plane. They interpret this decrease in metallicity as the transition from a thin disc to a thick disc dominated population, as opposed to any intrinsic thick disc gradient. Peng et al. (2013) also claim that there is no notable intrinsic vertical gradient in the thick disc. Conversely, there are studies which claim the existence of significant vertical gradients for this disc (Ruchti et al. 2011). Chen et al. (2011) measured the vertical gradient of the Milky Way using SDSS data. With fits over the range $1<|z|<3$ kpc they suggest a typical gradient of $-0.23\pm0.07$ dex kpc$^{-1}$. Unlike Kordopatis et al. (2013), Chen et al. (2011) attributed this gradient to be intrinsic to the thick disc. As discussed below, when we separate thin and thick discs based upon observationally-motivated vertical slices in $|z|$, we are unable to draw conclusions as to whether the thick disc has an intrinsic gradient or whether we are seeing the transition from thin to thick disc.

Recently, in Pilkington et al. (2012a) and Gibson et al. (2013) we have focused our work on how the radial gradients of Milky Way-scale simulations change as the galaxy evolves. Here, we extend that work to include how the radial metallicity gradient changes as a function of height above the plane (e.g. Bond et al. 2010; Carrell et al. 2012; Cheng et al. 2012) and how the vertical metallicity gradient changes as a function of radius (e.g. Carrell et al. 2012). We wish to test the efficacy of our various energy feedback schemes in recovering the inverted metallicity gradients claimed for thick disc stars of the Milky Way, and where having done so, identify *why* these simulations are successful in doing so. This does not constitute *sufficient* proof of a direct analogy to the Milky Way, but is a *necessary* condition for any putative picture for the formation and evolution of the thick disc, as our galaxies are not one-to-one Milky Way models, but disc galaxies of a similar mass and environment.

## 2. Simulations

The codes and specific simulations employed here have been detailed in Stinson et al. (2010) (MUGS: McMaster Unbiased Galaxy Simulations), Brook et al. (2012) (MaGICC: Making Galaxies In a Cosmological Context) and Few et al. (2014) (`109-CH`). Here, we only provide a summary of their primary characteristics.

Four of our simulations were realised with the Smoothed Particle Hydrodynamics (SPH) code Gasoline (Wadsley et al. 2004). Two variants of what have been referred to in our earlier papers as galaxies `g1536` and `g15784` were generated, one (MUGS) using the more conservative feedback scheme described by Stinson et al. (2010), and one (MaGICC) using the more energetic approach outlined by Brook et al. (2012). After Gibson et al. (2013), we refer to these four simulations as MUGS-g1536, MUGS-g15784, MaGICC-g1536 and MaGICC-g15784. The primary characteristics of each of the simulations used in our work are listed in Table 1.

Within Gasoline, stars can form when gas has become sufficiently cool (MUGS: $T_{\max}$<15000 K; MaGICC: $T_{\max}$<10000 K) and sufficiently dense (MUGS: $n_{\rm th}$>0.1 cm$^{-3}$; MaGICC: $n_{\rm th}$>9.3 cm$^{-3}$). When a gas particle is eligible for star formation, stars form according to: d$M_\star$/d$t$= $c_\star M_{\rm gas}/t_{\rm dyn}$, where $M_\star$ is the mass of stars formed in time d$t$, $M_{\rm gas}$ is the mass of a gas particle, $t_{\rm dyn}$ is the dynamical time of gas particles, and $c_\star$ is the star formation efficiency - i.e., the fraction of gas that will be converted into stars.

Supernova feedback follows the blastwave model of Stinson et al. (2006) with thermal energy (MUGS: $0.4 \times 10^{51}$ erg; MaGICC: $10^{51}$ erg) deposited to the surrounding ISM from each supernova. Cooling is disabled in the blast region (∼100 pc) for ∼10 Myr. All the simulations include heating from a uniform UV ionising background radiation field (Haardt & Madau 1996). Cooling within the simulations takes into account both primordial gas and metals. The metal cooling grid is derived using CLOUDY (v.07.02: Ferland et al. 1998) and is described in detail in Shen et al. (2010).

The MaGICC simulations also include radiation feedback from massive stars (Hopkins et al. 2011). While a typical massive star might emit ∼10$^{53}$ erg of radiation energy during its pre-SN lifetime, these photons do not couple efficiently to the surrounding ISM; as such, we only inject 10% of this energy in the form of thermal energy into the surrounding gas, and cooling is not disabled for this form of energy input. Of this injected energy, typically 90-100% is radiated away within a single dynamical time.

Raiteri et al. (1996) give a full description of the chemical evolution prescription in Gasoline; here we summarise the main points. In the MUGS runs only oxygen and iron were tracked, and the overall metallicity (Z) was assumed to be Z≡O+Fe, which led to an underprediction in Z of ∼0.2 dex. However, as described in Pilkington et al. (2012a), this does not affect the metallicity gradients. The MaGICC runs track seven elements from SNe and AGB stars, and assume Z≡O+Fe+C+N+Ne+Mg+Si. Metal diffusion is included (in both MUGS and MaGICC), such that unresolved turbulent mixing is treated as a shear-dependent diffusion term (Shen et al. 2010). This allows proximate gas particles to mix their metals. Metal cooling is calculated based on the diffused metals.

Simulation `109-CH` was run with the Adaptive Mesh Refinement (AMR) code ramses (Teyssier 2002), extended with our chemical evolution patch (ramses-ch - Few et al. 2012a and Few et al. 2014). ramses-ch tracks eight elements (H, C, N, O, Mg, Ne, Si, Fe) and the global metallicity Z. The gas density threshold for star formation is 0.3 cm$^{-3}$; once gas cells are eligible to form stars, they do so according to $c_\star \rho_{\rm gas}/t_{\rm ff}$, where $c_\star$ is the star formation efficiency, $\rho_{\rm gas}$ is the gas density, and $t_{\rm ff}$ is the local free-fall time of the gas.[1]

Feedback from SNe II events is presumed to be kinetic, while that from SNe Ia and AGB stars is thermal. Kinetic feedback deposits density, momentum, energy, and metals into all gas cells within a radial sphere of radius equivalent to two grid cells. The energy from each of these SNe II events is set to be $E_{\rm SN}=10^{51}$ erg. Kinetic feedback energy is determined by integrating the IMF for each stellar particle, and the momentum imparted to the gas depends then on the mass of the ejected material with an additional amount of swept-up material equal to the ejected gas mass times a factor $f_{\rm w}$ ($f_{\rm w}$=10 for `109-CH` - see Few et al. 2014). The thermal feedback is much simpler and only spread over the gas cell in which the star particle is located. The thermal energy feedback from SNe Ia assumes $E_{\rm SN}=10^{51}$ ergs; the physical scale over which this energy is deposited has an impact on the slope of the resulting metallicity gradient (Pilkington et al. 2012a; Gibson et al. 2013).

In ramses, cooling is computed assuming photoionisation equilibrium (with a uniform UV background; Haardt & Madau 1996) as a function of temperature for different metallicities and densities. The metal cooling grid is derived from calculations from the CLOUDY code (Ferland et al. 1998). Gas that is colder than $10^4$ K is cooled with metal fine-structure rates from Rosen & Bregman (1995).

## 3. Results

We focus our analysis on the disc stars associated with each of the five simulations described in §2. To isolate disc stars from spheroid stars, we employ the Abadi et al. (2003a) kinematic decomposition methodology, as used in earlier papers in this series. We decompose the full sample in two sub-populations according to the probability distribution function of $J_{\rm z}/J_{\rm circ}$: we take the distribution to the negative side and its mirror distribution as the spheroid stars and the rest constitute the disc stars, which all have positive values of $J_{\rm z}/J_{\rm circ}$. That way we reduce the bias because we remove the hottest population, most of them spheroid stars. When taking all the stars without doing any kinematical decomposition the results change ∼1% and the trends remain the same. Further, while our simulations are 'Milky Way-like', they should not be construed as being *identical* to the Milky Way or thought of as direct Milky Way models; any such analysis will be necessarily qualitative. We believe that it is the trends that are important here, rather than the absolute values. For example, while the vertical scaleheights of our 'thick discs' are not dissimilar to that of the Milky Way (∼1-1.5 kpc - Gilmore & Reid 1983), the radial scalelengths are ∼1.5-2× longer (cf. Bovy & Rix 2013). Even though the analysis which follows isolates radial 'regions' in the disc by 'kpc', rather than 'disc scalelength', it is important to emphasise that our results are robust and not contingent upon the selected radial range.

---

[1] ramses-ch uses the free-fall time to infer the star formation rate, while Gasoline uses the dynamical time; ramses-ch also does not employ an explicit variable called $c_\star$, but we have calculated its equivalent to frame the discussion here.

| Galaxy | IMF | $c_\star$ | $\epsilon$SNe | SR | $T_{\max}$ | $n_{\rm th}$ | $M$ |
|---|---|---|---|---|---|---|---|
| MUGS-g1536 | Kroupa | 0.05 | 40% | 0% | 15000 | 0.1 | 7.0 |
| MUGS-g15784 | Kroupa | 0.05 | 40% | 0% | 15000 | 0.1 | 14.0 |
| MaGICC-g1536 | Chabrier | 0.1 | 100% | 10% | 10000 | 9.3 | 6.8 |
| MaGICC-g15784 | Chabrier | 0.1 | 100% | 10% | 10000 | 9.3 | 14.3 |
| 109-CH | Kroupa01 | 0.01 | 100% | – | – | 0.3 | 7.1 |

**Table 1.** Primary characteristics of the simulations analysed in this work. Information for each column: 1. Simulation label; 2. Initial Mass Function (Kroupa≡Kroupa et al. 1993; Kroupa01≡Kroupa 2001; Chabrier≡Chabrier 2003); 3. Star formation efficiency; 4. Thermalised SNe energy fraction coupled to the interstellar medium (ISM); 5. Thermalised massive star radiation energy fraction coupled to the ISM; 6. Maximum allowable gas temperature for star formation (K); 7. Minimum gas density needed for star formation (cm$^{-3}$); 8. Total mass of the galaxy ($10^{11}$ M$_\odot$).

### 3.1. Radial Metallicity Gradients

Radial abundance gradients within the MUGS and MaGICC simulations were the focus of Pilkington et al. (2012a) and Gibson et al. (2013), with their emphasis being the redshift evolution of the gas-phase abundance gradients. Both works demonstrated the powerful diagnostic that gradients can play in constraining the nature and physical extent of feedback. Here, we move beyond the gas-phase *in-situ* radial abundance gradients and examine the impact on gradients when transitioning from the thin to the thick disc, contrasting with the behaviour seen in recent observational work and other simulations (e.g. Cheng et al. 2012; Carrell et al. 2012).

#### 3.1.1. [Fe/H] gradients

We begin looking at the distribution of the metallicity of stars with different ages. Figure 1 includes all disc stars in the radial range 5-10 kpc for one of our simulated galaxies, MaGICC-g15784 (the trend is very similar for the rest of them). The inner 5 kpc was avoided to filter out any remaining spheroid stars that were not removed by the kinematic decomposition. We also limit our sample to stars with $|z|<$3 kpc. As expected, the mean metallicity increases very rapidly during the first ∼4 Gyr of time and after that it remains almost constant.

Now we are ready to have a look at the gradients. In Fig. 2, we show the radial metallicity gradients (weighted by mass) for our five Milky Way-scale simulations, as a function of absolute height above the plane, including all disc stars in the radial range 5<$r$<10 kpc (we choose a similar radial range as that used in most studies in the literature). In the figure, data from observations is represented by lines and results from simulations with symbols, as noted in the inset. Although the radial ranges employed by the rest of the results are not always exactly the same as ours, it does not impact upon the results. We have checked the influence of changing the radial and vertical bins by ±2 kpc and the trends remain the same, even if the exact numbers vary somewhat. All the values shown in the plot are representative of the mid-point of each vertical height bin in every simulation and observation; they have not been weighted by the density distribution. This does not introduce any dramatic bias because all the data has been analysed in the same way.

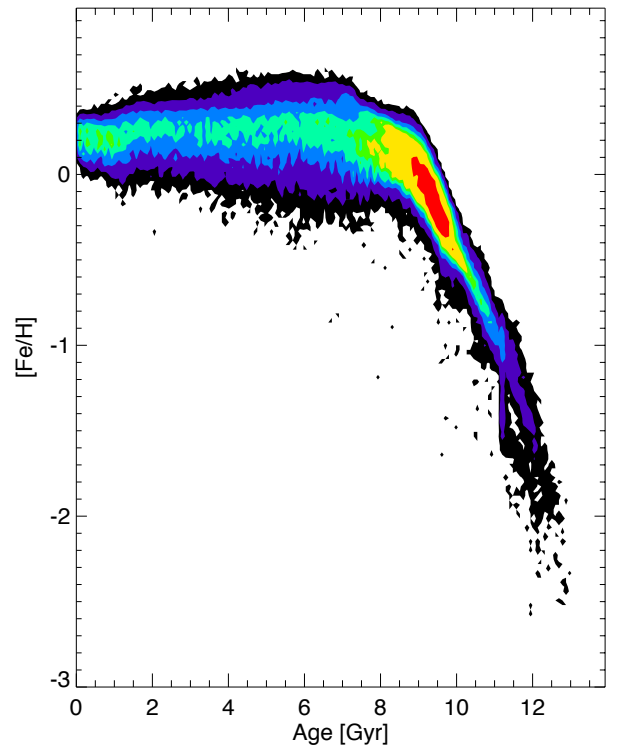

**Fig. 1.** Age-metallicity relation for the disc stars which fulfill the conditions 5<$r$<10 kpc and $|z|<$3 kpc. Old stars have a low value of [Fe/H], but it rapidly increases due to the SN Ia enrichment of the ISM. This figure corresponds to the galaxy MaGICC-g15784, but the relation is analogous for the rest of simulations in our sample.

From Fig. 2, one can see that (as expected), the gradients are negative at low scaleheights, but gradually flatten when moving to greater heights above the plane. The gradient within the MaGICC runs, with their substantially greater energy feedback, actually inverts (becoming positive) at heights ≳1 kpc. This transition occurs at heights which are also thought of as the transition between the thin and thick stellar discs of the Milky Way, in terms of the number counts of the latter beginning to dominate over the former (Jurić et al. 2008). Simulations which distribute the energy from supernovae over larger spatial scales, such as 109-CH, regardless of whether that feedback is 'conventional' or 'strong' energy-wise, also show gradient inversions when reaching heights associated with the classical thick disc. Conversely, the more conservative approaches to feedback, such as those applied to our MUGS realisations, while flattening with scaleheight, do not invert. The values of the radial gradients for the simulations employed here are listed in Table 2.

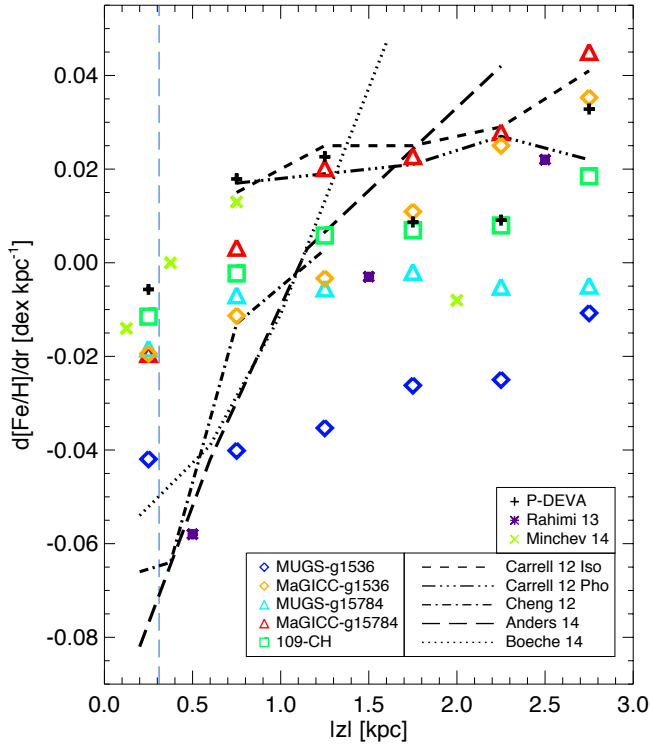

**Fig. 2.** Radial abundance gradient as a function of height from the plane. The figure includes our five simulations, MUGS-g1536 (in blue diamonds), MaGICC-g1536 (orange diamonds), MUGS-g15784 (cyan triangles), MaGICC-g15784 (red triangles) and 109-CH (green squares), for all disc stars in the radial range $5<r<10$ kpc. In addition, we present recent observational data and results from other simulations, drawn from the literature, overplotted in the panel (with the caveat that their radial and vertical selection criteria do not exactly match ours). Simulated results are plotted with symbols and observations with lines. Simulations from Rahimi et al. (2013) are represented with purple asterisks and the ones from Minchev et al. (2014) with light green crosses. The values for P-DEVA are indicated with black plus signs. Observational data from Carrell et al. (2012) (short dashed and triple dot dashed lines), Cheng et al. (2012) (dot dashed line), Anders et al. (2014) (long dashed line) and Boeche et al. (2014) (dotted line), are shown. The blue vertical dashed line corresponds to the effective force resolution of the simulations, i.e. 314 pc.

The inversion of the radial metallicity gradient with height is also suggested by observations of the Milky Way disc. In Fig. 2 we have included some of these studies, each of which typically possesses a slope uncertainty of $\sim$0.005 dex kpc$^{-1}$: Cheng et al. (2012) checked the radial metallicity gradient of old main-sequence turnoff stars from the SEGUE survey, in a radial range from 6 to 16 kpc; Carrell et al. (2012) studied a sample of F, G and K dwarf stars from SDSS DR8, through two different distance measurements - isochrone and photometric distances (their radial range extends from 7 to 10.5 kpc); Anders et al. (2014) explored the APOGEE data, investigating the red giant sample with $6<r<11$ kpc. Each of these studies finds trends that are reflected in our simulations.

Several complementary works to ours have also examined this behaviour, and in Fig. 2 we therefore incorporate the prediction for three of these models drawn from the literature. Each of them was realised using different methodologies and simulations from those in our extended suite: Minchev et al. (2014) employed a chemodynamical model to study the radial gradients between $5<r<10$ kpc and Rahimi et al. (2013) searched between 7 and 10 kpc in radii through their cosmological simulation. We also include one realisation of a Milky Way-like system run using the P-DEVA SPH code (Martínez-Serrano et al. 2008, 2009). It is interesting to note that in the Minchev et al. (2014) simulation, at a distance from the mid-plane greater than $\sim$2 kpc, the gradient reverts from 'inverted' to 'negative' again. This is related to the flaring of the younger disc in this run; this flaring is not sufficient at that scaleheight to allow a population of younger, more metal-rich, stars to populate that region, which is ultimately the reason for the gradient reverting back to negative.

There are more studies checking the radial metallicity gradients. For example, values from Boeche et al. (2013) agree well with our simulation results apart from stars with $|z|>0.8$ kpc where they find a more positive gradient (0.056 dex kpc$^{-1}$). The small number of stars in their sample though (middle panel of their figure 3) makes this inferred steep positive gradient somewhat uncertain.

Other trends for MUGS and MaGICC galaxies have previously been contrasted (e.g. Gibson et al. 2013; Obreja et al. 2013; Domínguez-Tenreiro et al. 2014). In all these cases, the consistency of our sets of galaxies with observational data is very satisfactory. The underlying physical processes responsible for the properties of these galaxies look the same for both the P-DEVA and the GASOLINE sets, including the origin of the spheroidal stellar components versus the disc ones (Domínguez-Tenreiro et al. 2015).

Because defining a thick disc can be done either 'spatially', as we have carried out by default (to allow for comparison with external edge-on discs), or 'chemically', we now check that our conclusions are robust to that thick disc 'definition'. In order to do so, we show the relation between [O/Fe] and stellar age for one of our simulated galaxies (MaGICC-g15784), in Fig. 3. Although not incorporated in this manuscript, the trends are equivalent for the other simulations. The normalisation to solar abundances has been taken from Anders & Grevesse (1989). Except for a shift to lower [O/Fe] values in the model, we can appreciate the similarity of this figure with observational works, such as figure 6 of Haywood et al. (2013), who present the same relation for a sample of solar neighbourhood stars (see also Snaith et al. 2015). We define our chemically selected thick disc as the stars which fulfill the condition [O/Fe]>$-0.05$ dex and thin disc the stars with [O/Fe]<$-0.05$ dex, following the two main peaks or populations identified in Fig. 3.

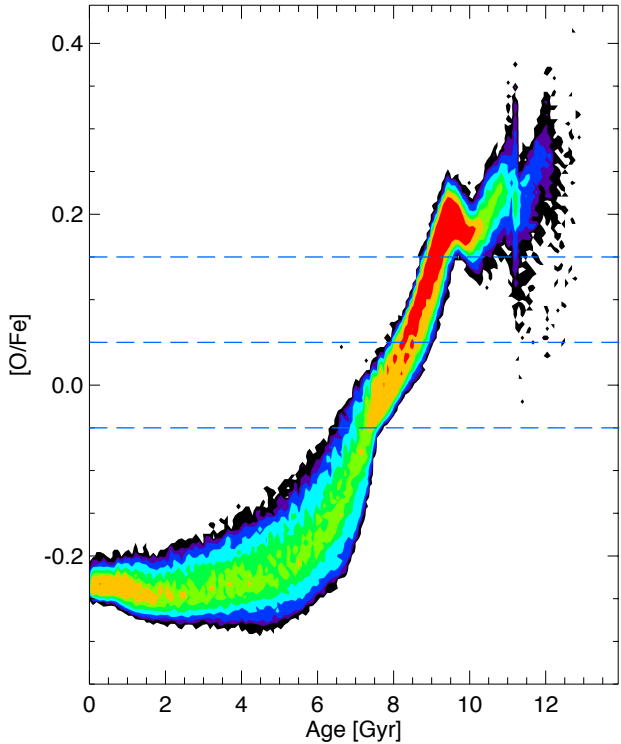
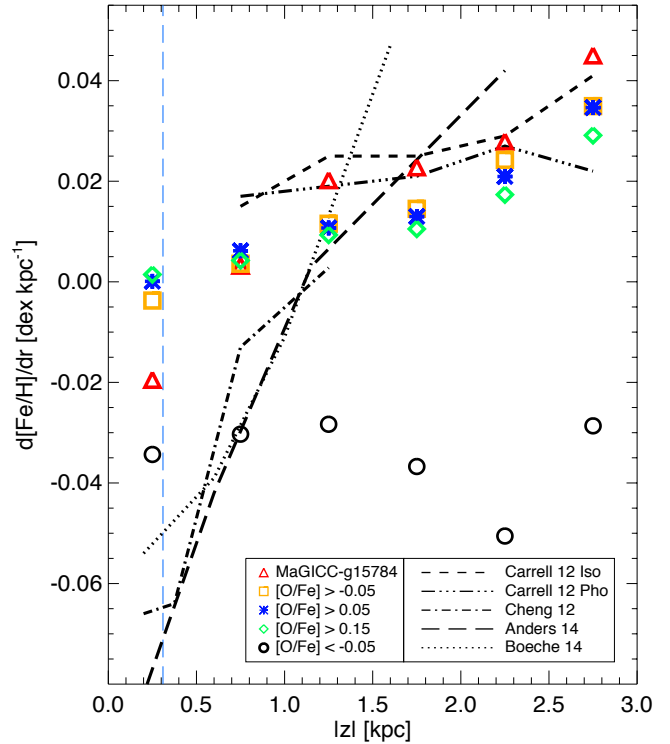

**Fig. 3.** [O/Fe] vs. age distribution for all disc stars included in Fig. 2 ($5<r<10$ kpc and $|z|<3$ kpc), for the galaxy MaGICC-g15784. The densest (i.e., most populated) regions are shown in red and the least dense in black. Blue horizontal dashed lines indicate the high-$\alpha$ cuts that are used in Fig. 4 to investigate the radial gradients of chemically selected populations, as opposed to spatially selected populations exposed in Fig. 2. We have two main peaks, with different age and [O/Fe], i.e. young $\alpha$-poor and old $\alpha$-rich stars. We call 'thick disc' to those disc stars with [O/Fe]$>-0.05$ dex and 'thin disc' to those with [O/Fe]$<-0.05$ dex.

**Fig. 4.** Radial [Fe/H] gradient vs. height inferred from isolating a thick disc via chemical characteristics, i.e. the high-$\alpha$ population. We repeat some data previously presented in Fig. 2: observational data and results from galaxy MaGICC-g15784, both indicated with the same linestyles and symbols as before. There also appear the gradients for the three high-$\alpha$ cuts displayed in Fig. 3 through the horizontal blue dashed lines ([O/Fe]$>-0.05$ dex in orange squares, [O/Fe]$>0.05$ dex in blue asterisks and [O/Fe]$>0.15$ dex in green diamonds) and the low-$\alpha$ cut ([O/Fe]$<-0.05$ dex in black circles).

We next return to the radial [Fe/H] gradients and their variation with height, but this time examining the effect produced by separating the thick disc via the selection of $\alpha$-enhanced stars (rather than simply a spatial cut). Figure 4 includes observational data and results from MaGICC-g15784, both with the same symbols and linestyles as in Fig. 2. Now we add the gradients for the sub-populations chemically separated according to the $\alpha$-cuts denoted by the blue horizontal lines in Fig. 3. We represent with orange squares and black circles the chemically defined thick and thin discs, respectively. The figure reveals the similarity of results when changing the thick disc selection criteria. The conclusion does not change: the thick disc has a positive radial metallicity gradient and it is negative for the thin disc. Table 3 contains the values of the gradients in the figure. In Fig. 4 we have also included more extreme thick disc conditions: blue asterisks show the result for subpopulation with [O/Fe]$>0.05$ dex and green diamonds for stars with [O/Fe]$>0.15$ dex. When increasing this high-$\alpha$ criteria, the contamination of the thin disc gets smaller and the gradients are even more positive (at low $|z|$ they are less and less negative) and flatter, which goes in agreement with our conclusions.

Increasing $|z|$ implies increasing age but not uniformly as a function of radius. While the outer disc is composed of stars with a range of ages at any height above the plane (due to flaring etc.), the inner disc is exclusively old at high $|z|$. This causes the inversion from negative to positive d[Fe/H]/d$r$, or positive to more positive as seen in Table 3 for an individual [O/Fe] cut. For any $|z|$ slice, increasing [O/Fe] results in flatter gradients because the higher the [O/Fe] cut, the smaller the age range available for the interplay of inside-out formation and disc flaring. Therefore, the gradients become flatter.

We have proved the consistency between our thick disc morphological selection and the chemical cut; indeed, our conclusions are unchanged by this choice. Whether $\alpha$-selected or spatially-selected, the radial metallicity gradients are positive for the thick disc and negative for the thin disc.

The phenomenon of the inversion of the radial metallicity gradient will be discussed in more detail in §4. Except for the next subsection concerning radial [O/Fe] gradients, in the rest of the paper we study the populations below and above 1 kpc separately, referring to them as thin and thick discs, respectively. We define the thick disc morphologically in contrast to a chemical/age cut because these $|z|$ slices allow better comparison with the Milky Way and external galaxies, as the latter are difficult to separate using detailed chemical abundances; such a spatial cut is a common means by which to separate thin and thick discs (e.g. Robin et al. 1996; Ojha 2001; Jurić et al. 2008; Hayden et al. 2015), but as stated earlier it is important to stress that our conclusions are robust to the choice of spatial vs. chemical definitions for the thick disc. Our approach is analogous to these aforementioned empirical thick disc works, and neatly avoids the ongoing debate regarding the use of the gap in the [α/Fe]-[Fe/H] plane to isolate disc components.

### 3.1.2. [O/Fe] gradients

The variation of the radial gradients in [O/Fe] as a function of $|z|$ is shown in Fig. 5 (the values are listed in Table 4). Here, we use [O/Fe] as a proxy for [α/Fe]. Lines are used to represent observational results and symbols for simulations. We do not see any clear differences between our five simulations, and, furthermore, most seem to agree with the observational work from Boeche et al. (2013). In that work, they measured the [α/Fe] radial gradient as a function of height from the galactic plane for their sample of dwarf stars. They defined the α-enhancement as [α/Fe]=([Mg/Fe]+[Si/Fe])/2. They carried out a similar analysis on the RAVE and GCS survey stars to check the consistency of their results. These two observations are shown by the dotted and triple dot dashed lines of Fig. 5, respectively. Both the RAVE and GCS results are somewhat limited by the scaleheights over which their data probes. Our simulations are also consistent with a number of other studies, including observations from APOGEE (Anders et al. 2014) and simulations from Minchev et al. (2014), which are also included in Fig. 5. The P-DEVA galaxy (Martínez-Serrano et al. 2008, 2009) also follows the same trends. Rahimi et al. (2013) found a steeper negative gradient at high $|z|$, although in the mid-plane their results are comparable to ours.

Gibson et al. (2013) presented the radial [O/Fe] gradients for the discs of MUGS-g1536 and MaGICC-g1536, showing both are relatively flat ($<-0.005$ dex kpc$^{-1}$). Pilkington & Gibson (2012) analysed how the [O/Fe] gradient of MUGS-g15784 evolves over time. Since redshift $z\sim 1$, stars have been born with relatively flat [O/Fe] gradients. At earlier times, the gradient was steeper, due to the presence of α-enhanced stars in the inner disc (a natural byproduct of the inside-out growth of the disc). At the present epoch these have now been distributed throughout the disc (due to various migration/heating/churning processes). This is similar to what was shown for [Fe/H], for all of the MUGS galaxies, in Pilkington et al. (2012b).

### 3.2. Vertical Metallicity Gradients

Vertical metallicity gradients have been found to be different for the thin and thick discs of the Milky Way (e.g. Marsakov & Borkova 2005, 2006; Soubiran et al. 2008). In Pilkington et al. (2012a), we studied the vertical metallicity gradients near the solar neighbourhood of the MUGS galaxies and found there to be no clear evidence for a two-component vertical structure in the density distribution. Given the force resolution (∼300 pc), this should not be surprising given that the thin disc's scaleheight is more or less only one softening length in the MUGS and MaGICC runs. Here, we expand upon our initial work, but only concentrate on the vertical gradients 1-3 kpc from the midplane (our so-called 'thick disc' stars), and how they vary as a function of galactocentric radius.

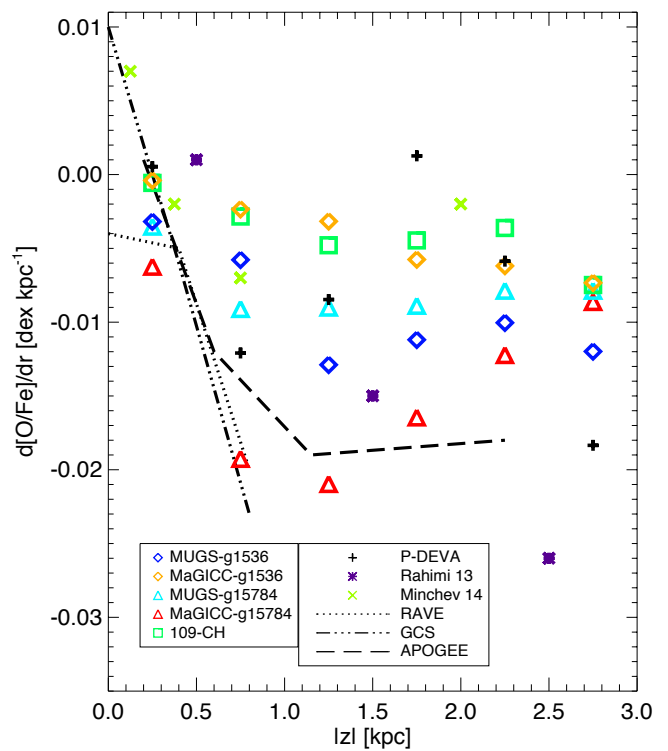

**Fig. 5.** [O/Fe] radial gradient as a function of distance from the galactic plane $|z|$. The disc stars with 5<$r$<10 kpc of our five simulations are plotted: MUGS-g1536 in blue diamonds, MaGICC-g1536 in orange diamonds, MUGS-g15784 in cyan triangles, MaGICC-g15784 in red triangles and `109-CH` in green squares. Results from some external simulations are also shown (P-DEVA in black plus signs, Rahimi et al. (2013) in purple asterisks and Minchev et al. (2014) in light green crosses), alongside the observational data from Boeche et al. (2013) (RAVE: dotted line; GCS survey: triple dot dashed line) and Anders et al. (2014) (APOGEE: dashed line).

The primary observational constraint is provided by Carrell et al. (2012), in which a strong negative gradient was inferred, through the thick disc, near the solar neighbourhood (with a vertical metallicity gradient of ∼−0.10 to −0.15 dex kpc$^{-1}$); such a steep gradient is entirely consistent with that inferred earlier by Marsakov & Borkova (2005). Figure 6 shows this information (which is also listed in Table 5) including our five simulations, the observational data from Carrell et al. (2012), and simulations from Rahimi et al. (2013) and the P-DEVA galaxy. Both observations and simulations have used the same vertical range to identify 'thick disc' stars: 1<$|z|$<3 kpc.

The major conclusion to take from Fig. 6 is that the simulations undertaken with enhanced and/or more spatially-distributed feedback (MaGICC and `109-CH`) each possess negative vertical metallicity gradients consistent with those observed in the solar neighbourhood of the Milky Way. A weak trend is also seen with galactocentric radius, the vertical gradients becoming shallower as one transitions from the inner to the outer disc. Conversely, the simulations with modest amounts of feedback (MUGS) show significantly flatter vertical gradients (at all galactocentric radii) than those observed in the Milky Way. We will return to the underlying physics driving these trends in §4.

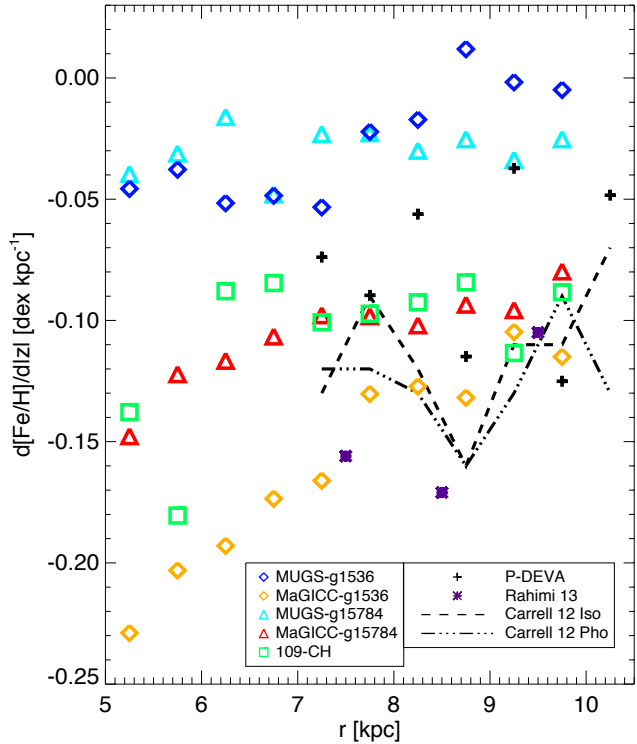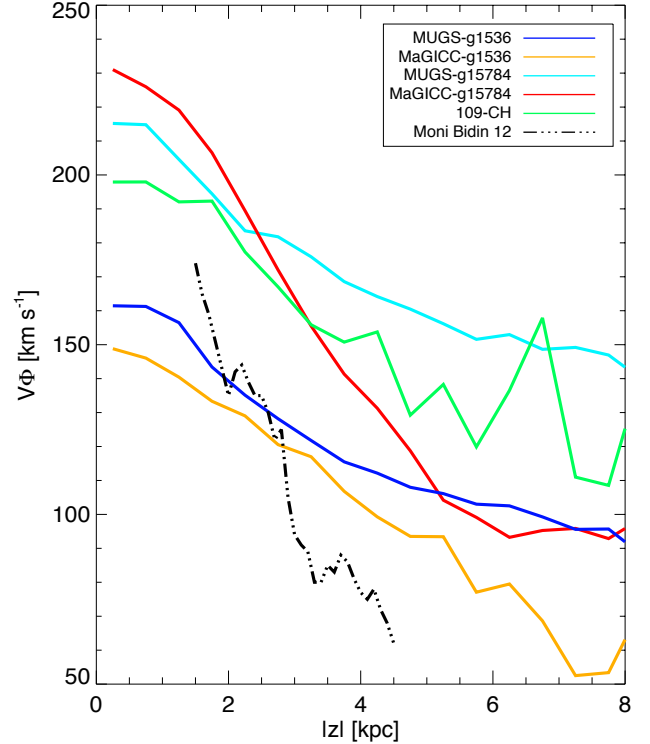

**Fig. 6.** Vertical metallicity gradient as a function of radius. The figure includes data from our five simulations: two MUGS galaxies, the same two galaxies but with MaGICC feedback conditions and `109-CH`. Observational data from Carrell et al. (2012) is shown with lines and results from the simulation in Rahimi et al. (2013) with purple asterisks. P-DEVA galaxy is indicated with black plus signs. All results are based upon stars selected from the vertical range 1<$|z|$<3 kpc.

**Fig. 7.** Rotational velocity of solar neighbourhood stars as a function of height above the plane. The four SPH simulations shown are MUGS-g1536 (blue line), MaGICC-g1536 (orange line), MUGS-g15784 (cyan line) and MaGICC-g15784 (red line), along with the AMR simulation `109-CH` (green line). Observational results from Moni Bidin et al. (2012) are also included (black triple dot dashed line).

### 3.3. Rotational Velocity Gradients

Recent observations (e.g. Bovy & Tremaine 2012) have examined the kinematic properties of stars in the solar neighbourhood as a function of height above the midplane. For example, Moni Bidin et al. (2012) found vertical gradients in the rotational velocity component of $\sim-30$ km s$^{-1}$ kpc$^{-1}$, in agreement with the work of Bond et al. (2010).

Figure 7 shows the rotational velocity $V_\phi$ as a function of height above the galactic plane for the analogous solar neighbourhoods of our suite of simulations. For each, this region was defined as an annulus spanning 7-10 kpc in galactocentric radius, and all stars assigned to the disc due to their kinematics were included in the derivation of the gradients. The change of radial range compared to the previous sections is done to make the comparison with observational results easier, as we have selected a range similar to theirs, but it should be noted that the results described here are not sensitive to the choice of the radial range. Four of the simulations (MUGS-g1536, MUGS-g15784, MaGICC-g1536 and `109-CH`) all show vertical gradients of $dV_\phi/d|z| \simeq -10$ km s$^{-1}$ kpc$^{-1}$. MaGICC-g15784 shows a much steeper vertical $V_\phi$ gradient of $\simeq -25$ km s$^{-1}$ kpc$^{-1}$, more in line with the observational findings for the Milky Way from Girard et al. (2006), Bond et al. (2010) and Moni Bidin et al. (2012).

The galaxies MUGS-g1536, MaGICC-g1536 and `109-CH` have a lower mass than the g15784 galaxies (Table 1). Even so, it is difficult to draw strong conclusions about the relation between the velocity gradient and mass, given the limited number of simulations in our sample. In addition, the effect of galaxy mass has not been well constrained observationally, due to limitations in resolving single stellar kinematics.

## 4. Discussion

Having shown the flattening of the radial metallicity gradient through the thick discs of our simulations, we now examine the mechanisms responsible for this behaviour. To do so, we will concentrate on MaGICC-g15784, as its abundance and kinematic profiles are the best match to the Milky Way.

We will examine the thick and thin disc stars separately following the $|z|$ division in §3, reminding the reader that, as we showed in Figs. 2 and 4, these two populations, when separated thusly, show opposite trends for the radial [Fe/H] gradients. Furthermore, we sub-divide the disc into three radial bins – 7<$r$<8 kpc, 10<$r$<11 kpc, and 14<$r$<15 kpc

– in order to have a sample of stars which populate the extrema of our radial bin in §3 and to extend the conclusions to larger radii.

### 4.1. Distribution Functions

In Figs. 8 and 9, we show the distribution functions of metallicity and formation radius for the stars associated with the aforementioned three radial bins, for two different vertical slices (corresponding to thick and thin discs).[2]

The [Fe/H] histograms in Fig. 8 are consistent with the radial metallicity gradients found in Fig. 2 (which were averaged at a given galactocentric radius). In Fig. 8, the top panel corresponds to the thick disc population while the bottom corresponds to the thin disc. The colour-coding in both panels represents the transition from inner disc (red), to middle (blue), to outer (green). The gradient in Fig. 2 has been determined from disc stars with galactocentric radii between 5-10 kpc. Beyond 10 kpc, the gradient actually starts to flatten. This is interesting because the chemically-defined thick disc more or less terminates at this radius, leaving a region defined by a single chemical component showing no significant radial metallicity gradient above the mid-plane. The thin disc, on the contrary, has a metallicity gradient which extends out to the furthest measured radii.

The migration of stars from their birth location, whether via bar-/arm-driven resonances, churning, or kinematic heating/diffusion will act to flatten metallicity gradients (e.g. Loebman et al. 2011; Pilkington & Gibson 2012; Pilkington et al. 2012a; Kubryk et al. 2013). In our study, we do not separate between these different mechanisms, in part because we do not capture the physics of resonances correctly due to lack of resolution. Here, we use the term 'radial migration' to refer to the change between birth and present day radius, without specifying which mechanism is responsible for causing the change; the mechanism, while interesting, is not critical for the specific analysis here.

To quantify the influence of migrators versus non-migrators, we refer to Fig. 9, which shows the distribution of formation radii for stars in the thick (upper panel) and thin (lower panel) discs. Transition from the red line, to the blue, and finally to the green corresponds to moving outwards radially (note: the current radial position is also indicated by the vertical dashed lines in the same colours). We can see that the inner thick disc is dominated by migrated/diffused stars, but when moving outwards the in-situ/locally-born population increases in relative importance. In the outermost bin (green line) one can define two different populations corresponding to each of the peaks in the distribution, where the inner peak would correspond to the migrators coming from the inner disc and the outer peak to the non-migrators/in-situ population.

Referring to the bottom panel of Fig. 9, one can see that for the thin disc there are two different populations at all current radii (again, spanning current radii of 7-15 kpc). The in-situ component is even more dominant, relatively speaking, than that seen for the thick disc (for each of the three radial bins), not surprisingly, given the inside-out growth of the thin disc.

---

[2] While not shown, negative age gradients for both the thick and thin discs exist, with no inversion encountered at higher scaleheights.

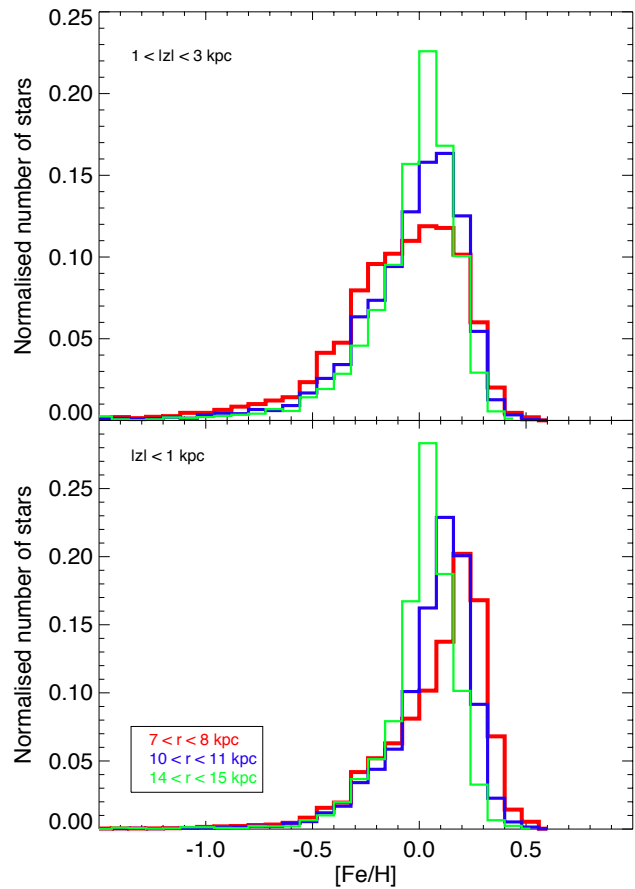

**Fig. 8.** Metallicity distribution functions for the so-called thick disc population ($1<|z|<3$ kpc, upper panel) and thin disc population ($|z|<1$ kpc, lower panel) associated with simulation MaGICC-g15784. Different colours correspond to different radial bins as indicated in the legend; when moving outwards we pass from red, to blue, to green.

### 4.2. Formation Radius

To explore the origin of the inverted metallicity gradients at high $|z|$, we build upon Fig. 9 and quantify the evolution of the stellar migration with time, metallicity and as a function of galactocentric radius. To relate this radial migration with the underlying metallicity gradients, in Fig. 10, we show the metallicity of the stars as a function of their formation radius for both thick (upper panels) and thin (lower panels) disc stars presently situated in the inner (left panels), middle (centre panels) and outer (right panels) disc of MaGICC-g15784; these present-day locations for said stars are denoted by the vertical dashed lines in each panel.

The solid lines in Fig. 10 reflect the mass-weighted radial [Fe/H] gradients of the young stars (a proxy for the instantaneous gas-phase abundances), with each colour representing a different epoch (red=past; purple=present-day). These act to illustrate the time evolution of the abundance gradients (using all disc stars within $|z|<3$ kpc and $5<r<15$ kpc). An earlier analogous version of this can be seen in figure 4 of Pilkington et al. (2012a), although in that case, they only examined MUGS-g15784 and Apollo (from the RaDES sample: Few et al. 2012b). The stronger feedback (with its enhanced redistribution of metals onto the largest scales) associated with the MaGICC galaxies (relative to MUGS or RaDES) acts to mitigate any significant evolution of the metallicity gradients, as identified by Gibson et al. (2013); we see that the slope of these lines in Fig. 10 does not change for different epochs in spite of the radial migration, as opposed to other studies.

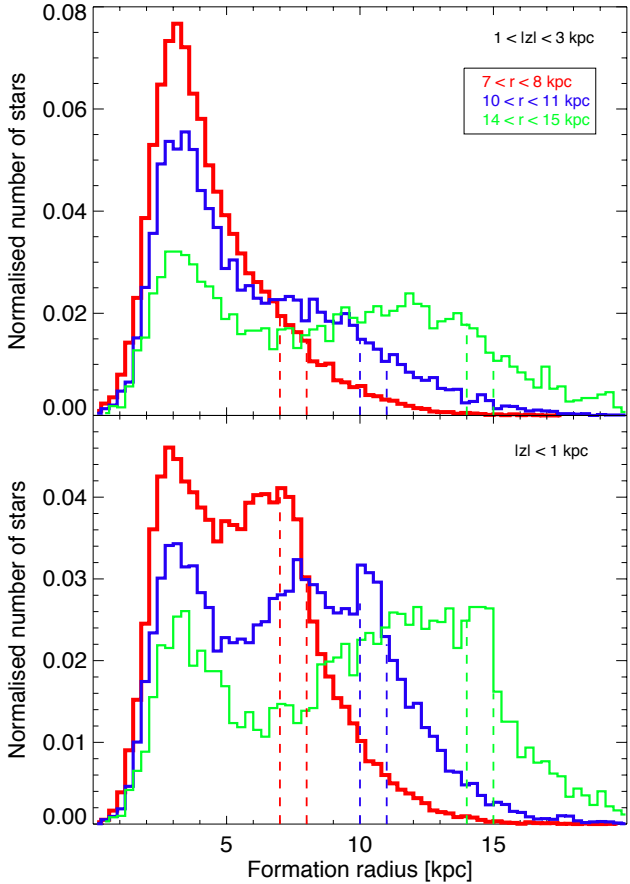

**Fig. 9.** Distribution of the radii of formation for thick and thin disc populations selected by $|z|$, in the top and bottom panels, respectively. Colour-coding corresponds to different present-day radial bins, as per the legend and Fig. 8. The vertical dashed lines show that current position of the stars in the associated radial bin (as indicated by their colour).

What is readily apparent from Fig. 10 is that the ratio of so-called 'in-situ' stars (those currently situated near to or within the vertical lines in each panel - i.e., those which have not moved significantly in a radial sense since birth) to those which have migrated a significant distance (corresponding primarily, but not exclusively, to the density enhancement seen in each panel near ∼3 kpc in formation radius) is higher in the outer regions for both the thick and thin discs. It is also readily apparent that the 'in-situ' population within the thin disc dominates at each radial bin, today, as to be expected according to the inside-out disc growth.

Specifically, referring to the inner thick disc (upper left panel of Fig. 10), its membership is formed primarily near a galactocentric radius of ∼3 kpc, $\gtrsim$10 Gyr ago (inferred from the age distribution for the density peak near ∼3 kpc), with metallicities $-0.3 \lesssim [Fe/H] \lesssim 0.0$. In contrast, the outer thick disc (upper right panel of Fig. 10) is also comprised of a tail of stars which formed nearby ('in-situ'), ∼1-5 Gyr ago, with metallicities $0.0 \lesssim [Fe/H] \lesssim +0.1$. Old stars are more centrally concentrated, which in turn is related to the disc inside-out formation.

In the thin disc, the situation is quite different. At all radii, the relatively metal-rich 'in-situ' component formed over the previous ∼7-8 Gyr dominates over any putative 'migration' component. That latter component exists at some level throughout the thin disc (which, for this simulation, is represented again by the density enhancement at relatively lower metallicities, originating near a galactocentric radius of ∼3 kpc), but its contrast with respect to the 'in-situ' component diminishes significantly with increasing radius. Unlike in the inner parts, the outermost region is quite similar for both thin and thick discs, which is telling us that thin disc stars can be found at high vertical distances; i.e. the disc of our simulated galaxy flares, consistent with Minchev et al. (2014).

We quantify the migration in Fig. 11, in which we compare the migration of stars drawn from each panel of Fig. 10. It is normalised to the average migration of each bin in the previously cited figure. Small values in the $x$-axis mean little travelling or migration. We can see that generally, in the three radial bins, thin disc stars have a greater ratio of in-situ stars than those in the thick disc. Although this quantity is very different when comparing the inner bin for the thick and thin discs (red and orange lines respectively), the difference reduces when moving to larger radial values, and finally, in our outermost radial bin the migration is quite similar for both discs (thick disc in green and thin disc in light green). Again, it is this 'contrast' that is the source of the inversion of the radial metallicity gradient in this simulation.

We make a consistency check and illustrate in Fig. 12 the same as Fig. 10 but separating the stars by their chemistry. Here, the top row shows the thick disc according to the abundance cut made in §3.1.1 and the bottom row the thin disc. The trends and conclusions are the same as in Fig. 10: the outer thick disc is composed of (by mean) younger, more metal-rich stars than its inner counterpart, which pull the radial metallicity gradient up, and invert it. Once again, we state that our conclusions are robust to the thick disc selection criteria.

Minchev et al. (2015) suggested that thick discs result from the imbedded flares of mono-age stellar populations, which should always result in discs growing inside-out. Such flaring would naturally give rise to the inversion of chemical gradients with increasing distance from the disc mid-plane, as we find here. Therefore, our results may be generic, given the similarity of results with Minchev et al. (2015), in spite of the completely different simulation technique and chemical enrichment models employed.

Bringing together the different arguments from this section, we can summarise by saying that the negative radial metallicity gradient seen in the thin disc can be inferred directly from tracing the metallicity of the density peak of the 'in-situ' population (bottom row of Fig. 10); that population dominates at all radii and reflects the inside-out nature of disc growth. Conversely, the positive radial metallicity gradient seen in the thick disc is driven by a 'contrast effect' between the 'in-situ' and 'migrator' populations; in moving from the inner thick disc, to the outer regions, the contribution from migrators drops dramatically, and the (relatively) more metal-rich in-situ population gradually flattens and, in some cases, inverts the gradient. Consequently, it is the tail of younger/in-situ stars (which is superimposed on a population of stars that have relocated from the centre) which shape the flattened/inverted gradients.

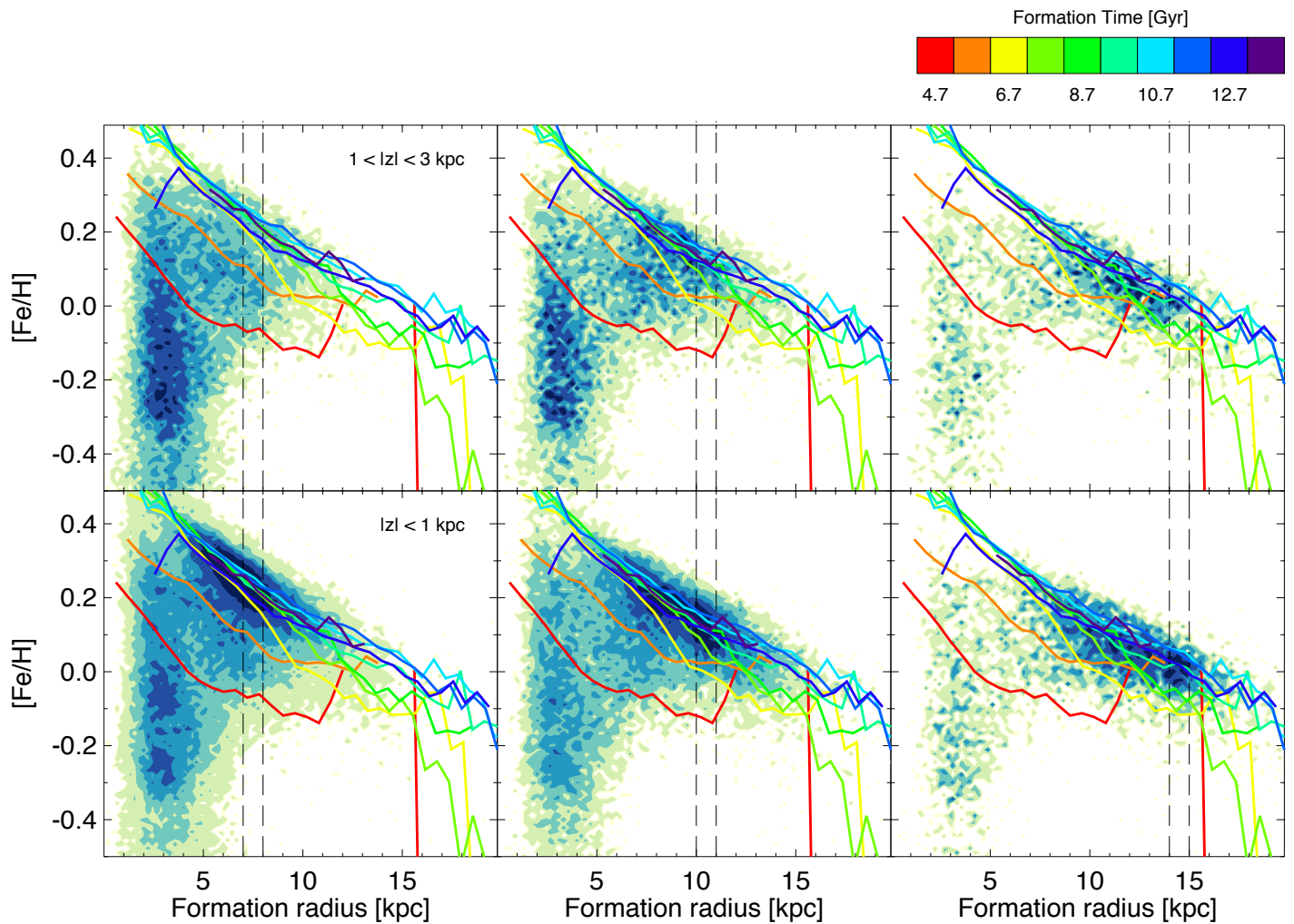

**Fig. 10.** Density plots of the metallicity for the thick (upper row) and thin (lower row) discs in terms of the formation radius of the stars are shown alongside the evolution of the metallicity gradient with time (solid lines), for galaxy MaGICC-g15784. Stars have been selected according to our spatial cut, as outlined in §3.1.1. The legend bar above the figure describes the difference in colour for the lines, which are the same in the six panels; each line indicates the behaviour of [Fe/H] with the formation radius for new born stars, at different time steps, as indicated in the colour bar. The density plots from left to right show the results for increasing current radii ranges, between $7 < r < 8$ kpc, $10 < r < 11$ kpc and $14 < r < 15$ kpc. The vertical dashed lines in each panel correspond to these radial bins.

## 5. Summary

This work has analysed five simulations, two with lower feedback from the MUGS suite, the same two galaxies but with higher feedback from the MaGICC suite, and one with the new AMR code RAMSES-CH (109-CH). The study presented in this paper complements that of Stinson et al. (2013), where they also looked at the formation of thin and thick discs, finding excellent agreement with observations from the Milky Way, when restricting the analysis to the examination of mono-abundance stellar populations.

We examine our suite of simulations in order to make an internal comparison between them and then judge the potential link to the Milky Way. Our simulations are slightly hotter, in a kinematic sense, than the Galaxy, so we restrict the analysis simply to an identification of the similarity in trends. To expand upon the work presented in Pilkington et al. (2012a) and Gibson et al. (2013) (who considered only radial gradients, and only in the mid-plane of the disc) this analysis has centered on:

1. The change of the radial metallicity gradient with height above the galactic plane. We find that MaGICC and 109-CH simulations, which include increased feedback energy (mixing metals more efficiently) and distribution of the energy from supernovae over larger spatial scales, respectively, show better agreement with observations from the Milky Way and other simulations. When mov-ing away from the plane of the disc, the gradient increases and becomes flatter in all simulations, but only the MaGICC and `109-CH` show it becoming inverted (it becomes positive, i.e. more metal-rich in the outskirts, relative to the inner parts of their respective discs, at $|z|\sim1$ kpc). We define the thick disc population as the one with positive radial gradient ($1<|z|<3$ kpc) and thin disc the one with $|z|<1$ kpc. We have chosen a spatial cut to separate thin and thick discs, as done observationally for extragalactic studies of thick discs, and for many Milky Way studies (e.g. Hayden et al. 2015). That said, we have confirmed that the conclusions are not impacted in a qualitative sense when applying a thick-thin disc discriminant based upon chemical arguments.

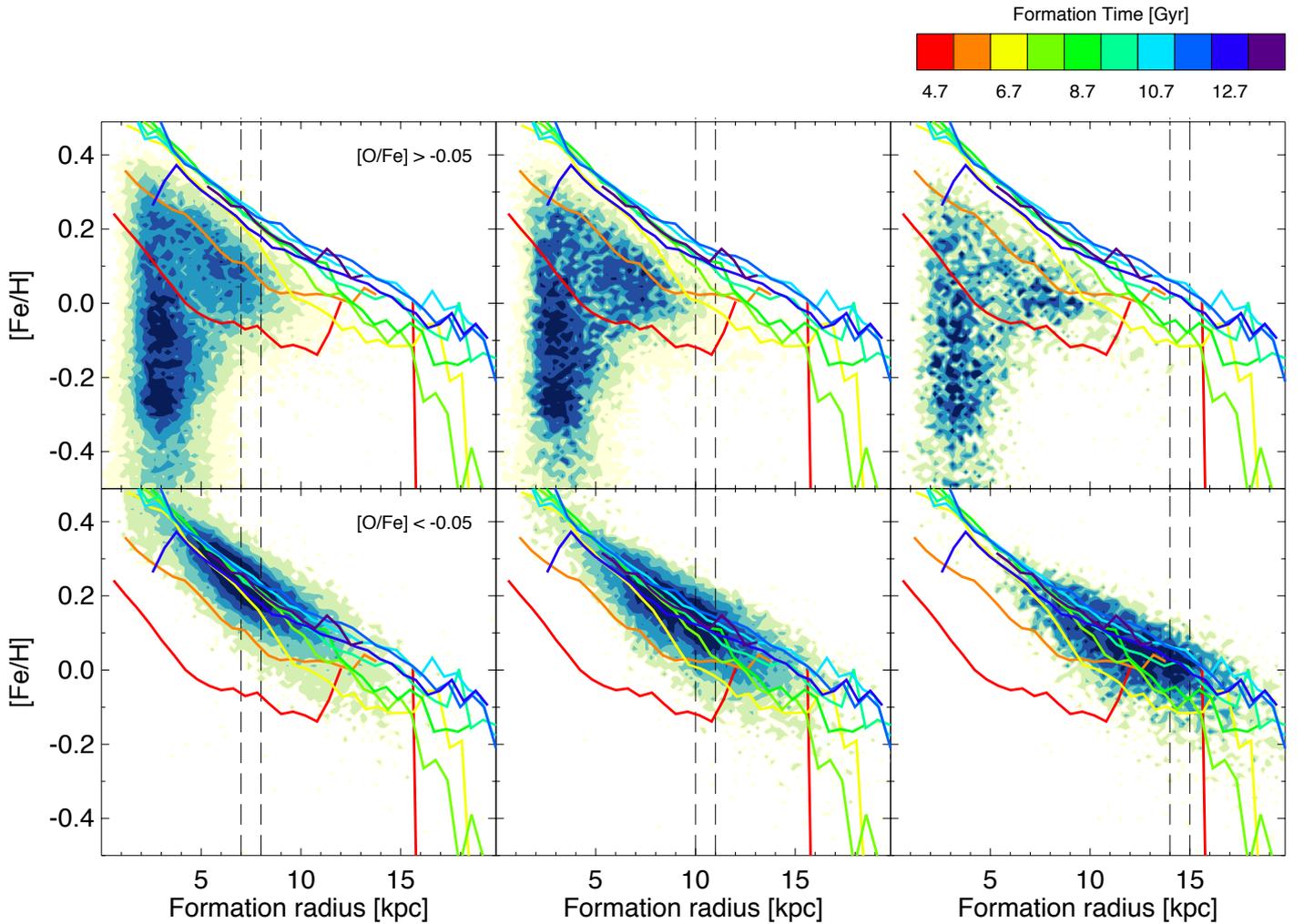

**Fig. 12.** Same as Fig. 10, but separating stars by abundance instead of morphologically. In this case the top and bottom rows indicate the chemically defined thick and thin discs, respectively. The thin disc does not contain very old stars, as we saw in Fig. 3. On the contrary, the thick disc is missing new-born stars.

2. The change of the vertical metallicity gradient as a function of radius, for the morphologically selected thick disc stars. The vertical gradients in all five simulations analysed in this paper show no dependence on radial distance from the galactic centre. The MaGICC and `109-CH` simulations both present vertical gradients of $\simeq-0.12$ dex kpc$^{-1}$, which agrees very well with the observational results from Carrell et al. (2012).

The MUGS galaxies exhibit shallower values. Observations of vertical gradients from Chen et al. (2011) and Kordopatis et al. (2013) both get vertical gradients in the thick disc with values $\simeq-0.2$ dex kpc$^{-1}$.

3. The change of the stellar rotational velocity at increasing $|z|$. One of our simulations (MaGICC-g15784) shows a vertical rotational velocity gradient in line with observations from Bond et al. (2010) and Moni Bidin et al. (2012) ($dV_\phi/d|z|\simeq-25$ km s$^{-1}$ kpc$^{-1}$). The other four simulations illustrate much shallower velocity gradients.

We next focused on the simulated galaxy which best reproduces the observations of the Milky Way thick disc – MaGICC-g15784 – and made a detailed study of its thin and thick disc stars (selected according to $|z|$) separately, at three different radial bins.

We analysed the distribution functions of [Fe/H] and formation radius. The metallicity histograms clearly reflect the trends noted before: when moving outwards from the centre of the galaxy, the metallicity of the thin disc decreases but in the thick disc it increases. From the histograms of the formation radius of the stars, we conclude that there is a significant influence from stars which migrate from their original location during their lifetimes, both in the thin and thick discs. We do not separate the different mechanisms here, simply referring to this process as 'migration', without any pre-disposition as to whether it is systematic 'churning' or a more random 'diffusion'.

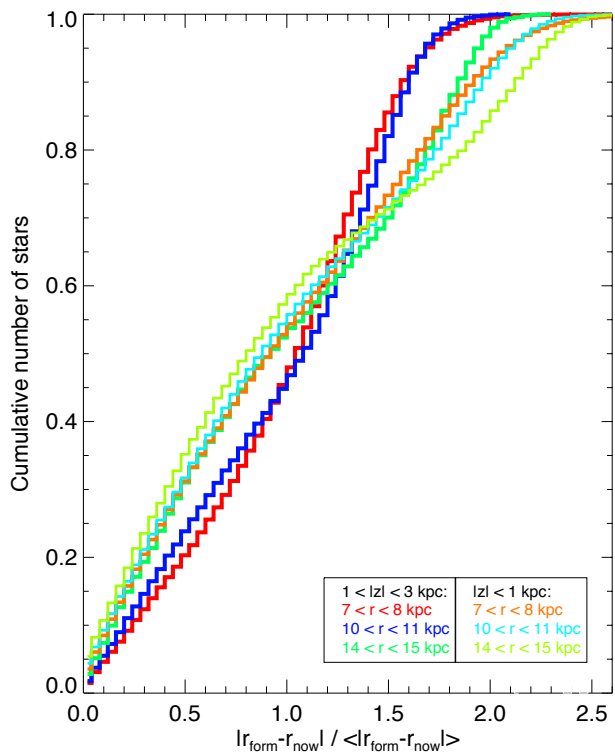

**Fig. 11.** Cumulative distribution function of the distance travelled by each star normalised to the average migration. Each line indicates the migration corresponding to one of the panels in Fig. 10. Large values along the $x$-axis represent greater degrees of migration. Darker colours correspond to the thick disc: stars which are currently located in the range $7<r<8$ kpc are shown in red, the ones at $10<r<11$ kpc in blue, and the ones at $14<r<15$ kpc in green. Lighter colours match the same radial bin but in the thin disc, with orange corresponding to our innermost bin ($7<r<8$ kpc), cyan the medium bin ($10<r<11$ kpc), and light green the outermost one ($14<r<15$ kpc).

To get a better understanding of the situation, in Fig. 10 we examined the relation between star formation history (by the time evolution of the metallicity gradients) and migration. This figure illustrates the situation for the thick and thin discs separately, for three radial bins. In the thick disc (top row), we see that from the left, to the middle and to the right panels, the centroid of the distribution moves from a very old low metallicity population to intermediate age metal-rich stars, making the metallicity gradient positive. In the thin disc (bottom row) the young population of in-situ, metal-rich stars dominates at small radii (left panel), increasing the mean metallicity and making the gradient negative.

The greatest difference between the two discs when referring to migration is that in the thin disc the ratio of in-situ stars to migrators is higher than in the thick disc. The contribution of the in-situ stars is what makes the radial metallicity gradient invert in MaGICC-g15784. The migrators erase all the trends because they form a larger population than the non-migrators (flattening the gradients in the outer galaxy), but still the in-situ population is what inverts the gradients in the thick disc. In other words, as shown in Fig. 11, the thin disc has a big influence of in-situ stars. The (spatially defined) outer thick disc has, as well, although these are also thin disc stars flaring upwards, which makes the thin and thick discs similar in terms of migration at our outermost radial bin. The innermost radial bin, however, is very different for each of the two disc components: the in-situ peak which dominates in the thin disc is missing for the thick one.

Figure 12 proves that our conclusions do not depend on the thick disc selection criteria. Both chemically and morphologically selected thick discs appear to have a positive radial metallicity gradient. Stars move radially and there is a frosting of in-situ star formation, with the latter dominating the further out in radius we go. Following the centroid of the points in the bottom row of any of these two figures one can confirm that the gradient is negative in the thin disc; similarly, following the centroid in the upper row it flattens and inverts. The difference between Fig. 10 and Fig. 12 lies in the presence, or lack thereof, of a tail of younger stars when doing the spatial cut.

Our simulations are somewhat kinematically hot, relative to the Milky Way (concluded from the proportion of stars which have migrated). From Fig. 10 we can make predictions concerning what might have happened had there been little or no migration (i.e., had the simulations been kinematically colder). In that case, the metallicity gradient would only be set by the temporal evolution of metallicity. In Fig. 10, stars would only 'travel' vertically as they would not change their radius. With a metallicity evolution such as that seen for MaGICC-g15784, the results would change quantitatively but not qualitatively; we would still see an inverted gradient in the thick disc and our conclusions would still be entirely valid. Put another way, the reason for the inversion of the metallicity gradient in our simulations is not stellar migration, but is driven by the stars that remain close to the radius at which they were born (the in-situ population).

According to our simulations, to reproduce the effects observed in the Galaxy we need to combine the high feedback employed in MaGICC and the contribution of the stars in the outer thick disc which have not migrated far from their birth location.

Our goal here has not been to undertake an internal comparison between different simulations, run with different codes and methodologies. Instead, we have aimed to provide a qualitative comparison of a broad suite of such simulations with the trends seen in the Milky Way, identify where the similarities lie, and then isolate the reasons for those similarities. The robustness of our conclusions are encouraging; larger and more precise datasets from Gaia, APOGEE, GALAH, WEAVE, 4MOST, etc., will prove invaluable in better constraining the underlying physical mechanisms we now believe responsible for the inverted gradients apparently intrinsic to the thick disc.

*Acknowledgements.* We acknowledge the insightful report provided by the referee, which improved dramatically the quality of the manuscript. MSM acknowledges the studentship support provided by the Jeremiah Horrocks Institute of the University of Central Lancashire. MSM also acknowledges EG for unconditional help and interesting discussions. BKG acknowledges the support of the UK's Sci-ence & Technology Facilities Council (ST/J001341/1). KP acknowledges the support of STFC through its PhD Studentship programme (ST/F007701/1). The generous allocation of resources from STFC's DiRAC Facility (COSMOS: Galactic Archaeology), the DEISA consortium, co-funded through EU FP6 project RI-031513 and the FP7 project RI-222919 (through the DEISA Extreme Computing Initiative), the PRACE-2IP Project (FP7 RI-283493), and CINES under the allocation 2010-c2011046642 made by GENCI (Grand Equipment National de Calcul Intensif). RS was supported by the Brain Korea 21 Plus Program (21A20131500002) and a Doyak Grant (2014003730). RS also acknowledges support from the EC through an ERC grant (StG-257720) and Fondecyt (3120135). JB acknowledges support for program HST-AR-12837 was provided by NASA through a grant from the Space Telescope Science Institute, which is operated by the Association of Universities for Research in Astronomy, Inc., under NASA contract NAS 5-26555.

|  | d[Fe/H] / d$r$ | | | | |
|---|---|---|---|---|---|
|  | MUGS-g1536 | MUGS-g15784 | MaGICC-g1536 | MaGICC-g15784 | 109-CH |
| $|z|$<0.5 kpc | −0.042 | −0.018 | −0.019 | −0.019 | −0.012 |
| 0.5<$|z|$<1.0 kpc | −0.040 | −0.007 | −0.011 | 0.003 | −0.002 |
| 1.0<$|z|$<1.5 kpc | −0.035 | −0.005 | −0.003 | 0.020 | 0.006 |
| 1.5<$|z|$<2.0 kpc | −0.026 | −0.002 | 0.011 | 0.023 | 0.007 |
| 2.0<$|z|$<2.5 kpc | −0.025 | −0.005 | 0.025 | 0.028 | 0.008 |
| 2.5<$|z|$<3.0 kpc | −0.011 | −0.005 | 0.035 | 0.045 | 0.018 |

**Table 2.** [Fe/H] radial gradients with increasing height for our five simulated Milky Way-like galaxies, measured in the radial range 5<$r$<10 kpc. The values are in dex kpc$^{-1}$.

|  | d[Fe/H] / d$r$ for MaGICC-g15784 | | | |
|---|---|---|---|---|
|  | [O/Fe]>−0.05 | [O/Fe]>0.05 | [O/Fe]>0.15 | [O/Fe]<−0.05 |
| $|z|$<0.5 kpc | −0.004 | 0.000 | 0.001 | −0.034 |
| 0.5<$|z|$<1.0 kpc | 0.004 | 0.006 | 0.004 | −0.030 |
| 1.0<$|z|$<1.5 kpc | 0.012 | 0.011 | 0.009 | −0.028 |
| 1.5<$|z|$<2.0 kpc | 0.015 | 0.013 | 0.011 | −0.037 |
| 2.0<$|z|$<2.5 kpc | 0.024 | 0.021 | 0.017 | −0.051 |
| 2.5<$|z|$<3.0 kpc | 0.035 | 0.035 | 0.029 | −0.029 |

**Table 3.** Same as Table 2, but considering sub-populations selected by [O/Fe], for galaxy MaGICC-g15784. [O/Fe] is given in dex and the gradients in dex kpc$^{-1}$. The four chemical conditions studied are: [O/Fe]>−0.05 dex, [O/Fe]>0.05 dex, [O/Fe]>0.15 dex, and [O/Fe]<−0.05 dex.

|  | d[O/Fe] / d$r$ | | | | |
|---|---|---|---|---|---|
|  | MUGS-g1536 | MUGS-g15784 | MaGICC-g1536 | MaGICC-g15784 | 109-CH |
| $|z|$<0.5 kpc | −0.003 | −0.004 | −0.000 | −0.006 | −0.001 |
| 0.5<$|z|$<1.0 kpc | −0.006 | −0.009 | −0.002 | −0.019 | −0.003 |
| 1.0<$|z|$<1.5 kpc | −0.013 | −0.009 | −0.003 | −0.021 | −0.005 |
| 1.5<$|z|$<2.0 kpc | −0.011 | −0.009 | −0.006 | −0.016 | −0.004 |
| 2.0<$|z|$<2.5 kpc | −0.010 | −0.008 | −0.006 | −0.012 | −0.004 |
| 2.5<$|z|$<3.0 kpc | −0.012 | −0.008 | −0.007 | −0.009 | −0.007 |

**Table 4.** [O/Fe] radial gradients (in dex kpc$^{-1}$) with increasing height for our five simulated Milky Way-like galaxies. The radial range used is 5<$r$<10 kpc.

## References

Abadi, M. G., Navarro, J. F., Steinmetz, M., & Eke, V. R. 2003a, ApJ, 591, 499
Abadi, M. G., Navarro, J. F., Steinmetz, M., & Eke, V. R. 2003b, ApJ, 597, 21
Afflerbach, A., Churchwell, E., & Werner, M. W. 1997, ApJ, 478, 190
Anders, E. & Grevesse, N. 1989, Geochimica et Cosmochimica Acta, 53
Anders, F., Chiappini, C., Santiago, B. X., et al. 2014, A&A, 564
Assmann, P., Fellhauer, M., Kroupa, P., Bruens, R. C., & Smith, R. 2011, MNRAS, 415, 1280
Bensby, T., Feltzing, S., & Lundström, I. 2003, A&A, 410, 527
Boeche, C., Siebert, A., Piffl, T., et al. 2014
Boeche, C., Siebert, A., Piffl, T., et al. 2013, A&A, 559
Bond, N. A., Ivezić, Ž., Sesar, B., et al. 2010, ApJ, 716, 1
Bournaud, F., Elmegreen, B. G., & Martig, M. 2009, ApJ, 707
Bovy, J. & Rix, H. W. 2013, ApJ, 779
Bovy, J., Rix, H.-W., & Hogg, D. W. 2012, ApJ, 751, 131
Bovy, J. & Tremaine, S. 2012, ApJ, 756, 89

|  | MUGS-g1536 | MUGS-g15784 | MaGICC-g1536 | MaGICC-g15784 | 109-CH |
|---|---|---|---|---|---|
| $5.0 < r < 5.5$ kpc | −0.046 | −0.040 | −0.229 | −0.148 | −0.138 |
| $5.5 < r < 6.0$ kpc | −0.038 | −0.031 | −0.203 | −0.122 | −0.180 |
| $6.0 < r < 6.5$ kpc | −0.052 | −0.016 | −0.193 | −0.117 | −0.088 |
| $6.5 < r < 7.0$ kpc | −0.049 | −0.048 | −0.173 | −0.107 | −0.085 |
| $7.0 < r < 7.5$ kpc | −0.053 | −0.023 | −0.166 | −0.098 | −0.101 |
| $7.5 < r < 8.0$ kpc | −0.022 | −0.023 | −0.130 | −0.098 | −0.097 |
| $8.0 < r < 8.5$ kpc | −0.017 | −0.030 | −0.127 | −0.102 | −0.093 |
| $8.5 < r < 9.0$ kpc | 0.012 | −0.025 | −0.132 | −0.093 | −0.084 |
| $9.0 < r < 9.5$ kpc | −0.002 | −0.034 | −0.105 | −0.096 | −0.113 |
| $9.5 < r < 10.0$ kpc | −0.005 | −0.025 | −0.115 | −0.080 | −0.088 |

Header: d[Fe/H] / d|z|

**Table 5.** [Fe/H] vertical gradients with increasing radius for our five simulated Milky Way-like galaxies. We have selected the disc stars with height $1 < |z| < 3$ kpc. The unit used for the gradients is dex kpc$^{-1}$.

Brook, C. B., Kawata, D., Gibson, B. K., & Freeman, K. C. 2004, ApJ, 612, 894
Brook, C. B., Stinson, G., Gibson, B. K., Wadsley, J., & Quinn, T. 2012, MNRAS, 424, 1275
Carollo, D., Beers, T. C., Chiba, M., et al. 2010, ApJ, 712, 692
Carrell, K., Chen, Y., & Zhao, G. 2012, AJ, 144, 185
Casetti-Dinescu, D. I., Girard, T. M., Korchagin, V. I., & van Altena, W. F. 2011, ApJ, 728, 7
Chabrier, G. 2003, PASP, 115, 763
Chen, Y. Q., Zhao, G., Carrell, K., & Zhao, J. K. 2011, AJ, 142, 184
Cheng, J. Y., Rockosi, C. M., Morrison, H. L., et al. 2012, ApJ, 746, 149
Comerón, S., Elmegreen, B. G., Knapen, J. H., et al. 2011, ApJ, 741
Domínguez-Tenreiro, R., Obreja, A., Brook, C. B., et al. 2015, ApJ, 800
Domínguez-Tenreiro, R., Obreja, A., Granato, G. L., et al. 2014, MNRAS, 439, 3868
Ferland, G. J., Korista, K. T., Verner, D. A., et al. 1998, PASP, 110, 761
Few, C. G., Courty, S., Gibson, B. K., et al. 2012a, MNRAS, 424, L11
Few, C. G., Courty, S., Gibson, B. K., Michel-Dansac, L., & Calura, F. 2014, MNRAS, 444, 3845
Few, C. G., Gibson, B. K., Courty, S., et al. 2012b, A&A, 547
Freeman, K. 2012, Structure and Evolution of the Milky Way, ed. A. Miglio, J. Montalbán, & A. Noels, 137
Gibson, B. K., Pilkington, K., Brook, C. B., Stinson, G. S., & Bailin, J. 2013, A&A, 554
Gilmore, G. & Reid, N. 1983, MNRAS, 202, 1025
Gilmore, G., Wyse, R. F. G., & Norris, J. E. 2002, ApJ, 574, L39
Girard, T. M., Korchagin, V. I., Casetti-Dinescu, D. I., et al. 2006, AJ, 132, 1768
Haardt, F. & Madau, P. 1996, ApJ, 461, 20
Hayden, M. R., Bovy, J., Holtzman, J. A., et al. 2015
Haywood, M., Di Matteo, P., Lehnert, M. D., Katz, D., & Gómez, A. 2013, A&A, 560
Hopkins, P. F., Quataert, E., & Murray, N. 2011, MNRAS, 417, 950
Ivezić, Ž., Sesar, B., Jurić, M., et al. 2008, ApJ, 684, 287
Jurić, M., Ivezić, Ž., Brooks, A., et al. 2008, ApJ, 673, 864
Kazantzidis, S., Bullock, J. S., Zentner, A. R., Kravtsov, A. V., & Moustakas, L. A. 2008, ApJ, 688, 254
Kordopatis, G., Hill, V., Irwin, M., et al. 2013, A&A, 555
Kroupa, P. 2001, MNRAS, 322, 231
Kroupa, P. 2002, MNRAS, 330, 707
Kroupa, P., Tout, C. A., & Gilmore, G. 1993, MNRAS, 262, 545
Kubryk, M., Prantzos, N., & Athanassoula, E. 2013, MNRAS, 436, 1479
Lee, Y. S., Beers, T. C., An, D., et al. 2011, ApJ, 738, 187
Loebman, S. R., Roskar, R., Debattista, V. P., et al. 2011, ApJ, 737
Luck, R. E., Andrievsky, S. M., Kovtyukh, V. V., Gieren, W., & Graczyk, D. 2011, AJ, 142, 51
Marsakov, V. A. & Borkova, T. V. 2005, Astronomy Letters, 31, 515
Marsakov, V. A. & Borkova, T. V. 2006, Astronomy Letters, 32, 376
Martínez-Serrano, F. J., Serna, A., Doménech-Moral, M., & Domínguez-Tenreiro, R. 2009, ApJ, 705
Martínez-Serrano, F. J., Serna, A., Domínguez-Tenreiro, R., & Mollá, M. 2008, MNRAS, 388, 39
Minchev, I., Chiappini, C., & Martig, M. 2014
Minchev, I., Famaey, B., Quillen, A. C., et al. 2012, A&A, 548
Minchev, I., Martig, M., Streich, D., et al. 2015, ApJ, 804
Moni Bidin, C., Carraro, G., & Méndez, R. A. 2012, ApJ, 747, 101
Navarro, J. F., Abadi, M. G., Venn, K. A., Freeman, K. C., & Anguiano, B. 2011, MNRAS, 89
Nemec, J. M. & Nemec, A. F. L. 1993, AJ, 105, 1455
Nordström, B., Mayor, M., Andersen, J., et al. 2004, A&A, 418, 989
Obreja, A., Domínguez-Tenreiro, R., Brook, C., et al. 2013, ApJ, 763
Ojha, D. K. 2001, MNRAS, 322, 426
Pasetto, S., Grebel, E. K., Zwitter, T., et al. 2012, A&A, 547, A71
Peng, X., Du, C., Wu, Z., Ma, J., & Zhou, X. 2013, MNRAS, 434, 3165
Pilkington, K., Few, C. G., Gibson, B. K., et al. 2012a, A&A, 540
Pilkington, K. & Gibson, B. K. 2012, in Astronomical Society of the Pacific Conference Series, Vol. 458, Galactic Archaeology: Near-Field Cosmology and the Formation of the Milky Way, ed. W. Aoki, M. Ishigaki, T. Suda, T. Tsujimoto, & N. Arimoto, 241
Pilkington, K., Gibson, B. K., Brook, C. B., et al. 2012b, MNRAS, 425, 969
Qu, Y., Di Matteo, P., Lehnert, M. D., & van Driel, W. 2011, A&A, 530
Quinn, T. & Binney, J. 1992, MNRAS, 255, 729
Rahimi, A., Carrell, K., & Kawata, D. 2013
Raiteri, C. M., Villata, M., & Navarro, J. F. 1996, Mem. Soc. Astron. Italiana, 67, 817
Reddy, B. E., Lambert, D. L., & Allende Prieto, C. 2006, MNRAS, 367, 1329
Rix, H.-W. & Bovy, J. 2013, A&A Rev., 21
Robin, A. C., Haywood, M., Creze, M., Ojha, D. K., & Bienayme, O. 1996, A&A, 305
Rosen, A. & Bregman, J. N. 1995, ApJ, 440
Ruchti, G. R., Fulbright, J. P., Wyse, R. F. G., et al. 2011, ApJ, 737
Schönrich, R. & Binney, J. 2009, MNRAS, 396, 203
Shaver, P. A., McGee, R. X., Newton, L. M., Danks, A. C., & Pottasch, S. R. 1983, MNRAS, 204, 53
Shen, S., Wadsley, J., & Stinson, G. 2010, MNRAS, 1043
Simpson, J. P., Colgan, S. W. J., Rubin, R. H., Erickson, E. F., & Haas, M. R. 1995, ApJ, 444, 721
Snaith, O. N., Bailin, J., Gibson, B. K., et al. 2015, MNRAS
Soubiran, C., Bienaymé, O., Mishenina, T. V., & Kovtyukh, V. V. 2008, A&A, 480, 91
Steinmetz, M., Zwitter, T., Siebert, A., et al. 2006, AJ, 132, 1645
Stinson, G., Seth, A., Katz, N., et al. 2006, MNRAS, 373, 1074
Stinson, G. S., Bailin, J., Couchman, H., et al. 2010, MNRAS, 408, 812
Stinson, G. S., Bovy, J., Rix, H.-W., et al. 2013, MNRAS, 436, 625
Teyssier, R. 2002, A&A, 385, 337
Vera-Ciro, C., D'Onghia, E., Navarro, J., & Abadi, M. 2014, ApJ, 794
Villalobos, Á. & Helmi, A. 2008, MNRAS, 391, 1806
Wadsley, J. W., Stadel, J., & Quinn, T. 2004, New Astronomy, 9, 137
Wyse, R. F. G., Gilmore, G., Norris, J. E., et al. 2006, ApJ, 639, L13
Yoachim, P. & Dalcanton, J. J. 2006, AJ, 131, 226